\documentclass[11pt,twoside]{article}
\usepackage{asp2010}
\resetcounters

\begin{document}

\title{Do sub-millimeter galaxy number counts provide evidence for a top-heavy IMF?}
\author{Christopher C. Hayward$^{1}$, Desika Narayanan$^{1}$, Patrik Jonsson$^{1}$, \\
T. J. Cox$^{2}$, Du\v{s}an Kere{\v s}$^{1}$, Philip F. Hopkins$^{3}$, and Lars Hernquist$^{1}$
\affil{$^1$Harvard-Smithsonian Center for Astrophysics, 60 Garden St., Cambridge, MA 02138, USA\\
$^2$Observatories of the Carnegie Institution of Washington, 813 Santa Barbara Street, Pasadena, CA 91101, USA\\
$^3$Department of Astronomy and Theoretical Astrophysics Center, University of California Berkeley, Berkeley, CA 94720, USA}}

\begin{abstract}
Theoretical models have had difficulty matching the observed number density of sub-millimeter galaxies (SMGs), causing
some authors \citep[e.g.,][]{Baugh:2005} to suggest that SMGs provide evidence for a top-heavy initial mass function (IMF).
To test this claim, we have, for the first time, combined high-resolution 3-D hydrodynamic simulations of isolated and merging massive,
gas-rich galaxies, radiative transfer, and a semi-empirical merger rate model to predict the number density of SMGs.
Our model can reproduce the observed SMG number density even when using a standard (Kroupa) IMF.  The agreement
is due to a combination of relatively long sub-mm duty cycles for mergers (a few times $10^{8}$ years for our most massive models),
which owe to our combination of high-resolution 3-D hydrodynamic simulations and dust radiative transfer;
sufficient number densities of massive, gas-rich mergers; and the decrease in sub-mm counts observed by recent deep/wide
surveys \citep[e.g.,][]{Austermann:2010} relative to previous surveys. Our results suggest that the observed SMG number counts
do not provide evidence for a top-heavy IMF at high redshift.
\end{abstract}

\section{Introduction}

Sub-millimeter galaxies (SMGs) are a population of extremely luminous
\citep[$L_{\rm bol} \sim 10^{12} - 10^{13} L_{\odot}$; e.g.,][]{Kovacs:2006}
high-redshift galaxies that emit almost all of their bolometric luminosity
in the IR \citep[for a review see ][]{Blain:2002}.
SMGs are primarily powered by star formation rather than AGN \citep[C. Hayward et al. 2010a, in prep.;][]{Alexander:2005},
with inferred star formation rates (SFRs) of order $10^2-10^3$ $M_{\odot}$ yr$^{-1}$.
Some models suggest that the stellar initial mass function (IMF) may be top-heavy
in such high-density environments (e.g., Mark Krumholz's proceeding in this volume), so SMGs are an excellent place to look for
IMF variation.

The sub-mm flux of a galaxy depends on the amount of energy
absorbed by dust and the mass, geometry, and composition of the dust. A top-heavy IMF means that more massive stars are formed
per unit SFR, resulting in greater luminosity emitted and dust mass produced per unit SFR. The
combination of increased luminosity heating the dust and increased dust mass leads to greater sub-mm flux. Therefore the
sub-mm flux of a galaxy -- and thus the number density of sub-mm sources -- predicted by a theoretical model will depend
sensitively on the assumed IMF.

To date, models have had difficulty reproducing the observed sub-mm number counts
using a standard IMF. The number counts predicted by the semi-analytic model (SAM) of \citet{Baugh:2005} 
were less than the observed counts by factors of $\gtrsim 20$. Of the model variations Baugh et al. tested,
only by assuming a flat IMF ($dN/d \log M = $ constant) in starbursts were they able to reproduced the observed counts.
This increased the sub-mm counts from starbursts by a factor of $\sim10^3$. However, the large parameter space of SAMs
often yields multiple qualitatively distinct solutions that satisfy all observational constraints \citep[e.g.,][]{Lu:2010},
leaving open the possibility that the SAM of Baugh et al. does not \emph{require} a top-heavy IMF to match the observed
counts.

As an alternate approach, \citet{Dave:2010} presented a large cosmological simulation for which they used the observed
SMG number density to calculate
how many of their simulated galaxies should be SMGs, assuming that the SMGs were the most highly star-forming galaxies
in the simulation. The galaxies they identified as SMGs had SFRs $\lesssim 3$ times those inferred from the IR
luminosities of SMGs. One possible explanation for the discrepancy is that the IMF in SMGs is top-heavy; if so, the star
formation rates inferred assuming a standard IMF would be higher than the true SFRs of the SMGs. (Note that the required
change to the IMF is much milder than that required by Baugh et al.)
However, there are multiple other potential explanations for the discrepancy (see Dav\'{e} et al. for discussion of these).
Furthermore, it is not clear if the resolution of the Dave et al. simulation is sufficient to fully resolve the SFR enhancement
caused by tidal torques in galaxy mergers. Finally, our models show that the simple rank ordering between SFR and sub-mm flux
assumed by Dav\'{e} et al. does not hold (C. Hayward et al. 2010b, in prep.), so the evidence provided by their work for a
top-heavy IMF is indirect at best.

The discrepancy between modeled and observed SMG number densities provides tantalizing but indirect evidence for a
top-heavy IMF. However, the complexity of the modeling leaves open the possibility that something other than a top-heavy IMF
can explain the discrepancy.
In this work we present a novel method to predict the sub-mm number counts:
we combine high-resolution hydrodynamic simulations of isolated and merging disk galaxies, 3-D radiative transfer (RT)
including a fully self-consistent calculation of the dust temperatures, and a semi-empirical model for merger rates.
Preliminary results show that our model can reproduce the observed sub-mm number densities
even when a standard IMF is used.

\section{Methodology} \label{S:methods}

In \citet{Narayanan:2010smg} we presented a merger-driven model for SMGs. We performed high-resolution {\sc Gadget-2}
\citep{Springel:2005gadget}
3-D hydrodynamic simulations of merging and isolated massive, gas-rich disk galaxies scaled to $z = 3$ and
used the 3-D adaptive grid polychromatic Monte Carlo dust RT code {\sc Sunrise} \citep{Jonsson:2006sunrise,Jonsson:2010sunrise}
to calculate the spectral energy distributions (SEDs) of the simulated mergers. The model successfully reproduced the
range of sub-mm fluxes, UV-mm SED, and bulge, black hole, and gas masses of SMGs. Furthermore, in \citet{Narayanan:2009}
we calculated the molecular line emission of the simulated galaxies using the 3-D non-LTE molecular line RT code
{\sc Turtlebeach} \citep{Narayanan:2008}, finding that our models were consistent with the observed line widths, compact CO spatial
extent, and excitation conditions of SMGs. Finally, our models naturally explain the intersection of the SMG and dust-obscured
galaxy (DOG) populations \citep{Narayanan:2010dog}. While the consistency between our models and
many observed properties of SMGs is encouraging, we have not yet shown
that our models can reproduce the observed SMG number counts, a key test of potential SMG models. Calculating the number
counts predicted by our models is the goal of the work described here.

Here we recapitulate the key details of the simulations presented in \citet{Narayanan:2010smg} and describe the primary difference
between our simulations and theirs.
The {\sc Gadget-2} hydrodynamic simulations include star formation via the volume-density-dependent
Kennicutt-Schmidt law \citep{Kennicutt:1998,Schmidt:1959}
\citep[consistent with observations of high-redshift galaxies; e.g.,][]{Narayanan:2010ks}, the two-phase ISM model of
\citet[][hereafter SH03]{Springel:2003},
and black hole accretion and feedback following \citet{Springel:2005feedback}. In post-processing we
use the dust RT code {\sc Sunrise} \citep{Jonsson:2006sunrise,
Jonsson:2010sunrise}
to calculate the UV-mm SEDs of the simulated galaxies assuming a \citet{Kroupa:2001} IMF. {\sc Sunrise} calculates the temperature
of each dust grain species assuming thermal equilibrium and self-consistently treats dust self-absorption, which is key for
an accurate determination of the sub-mm flux owing to the high optical depths of the central regions of the mergers.

The calculations presented here differ from those of \citet{Narayanan:2010smg} primarily in the treatment of the obscuration from cold
clouds in the ISM. \citet{Narayanan:2010smg} assumed that the young stars present in the simulations were completely
obscured by their molecular gas birth clouds for 10 Myr \citep[$f_{\rm PDR} = 1$ in the parlance of][]{Groves:2008}. The
attenuation and dust emission owing to these birth clouds was calculated via the sub-resolution model
of \citet{Groves:2008}. (See \citet{Jonsson:2010sunrise} for details of the {\sc Sunrise} implementation of the model.)
\citet{Narayanan:2010smg}
neglected any other obscuration from the cold phase of the SH03 two-phase ISM model, so that stellar radiation
only encountered dust in the birth cloud (if the radiation is from a star particle with age $< 10$ Myr) and the diffuse phase of the ISM.
Instead, we opt to neglect any sub-resolution obscuration from the birth clouds ($f_{\rm PDR} = 0$) and model the
high obscuration by ignoring the SH03 multiphase breakdown for the purposes of the RT calculation, assuming that
the total SPH gas density is distributed uniformly on sub-resolution
scales ($\lesssim$ 300 pc). By eliminating use of the SH03 sub-resolution model we can more effectively test how the assumptions
of our model affect the
predicted number counts. The consequences of this choice will be discussed in detail in C. Hayward et al. (2010c, in prep.).

In order to predict SMG number counts from our models, we combine sub-mm lightcurves from the {\sc Gadget}+{\sc Sunrise}
simulations with the semi-empirical merger rate
model of \citet{Hopkins:2010merger_rates}. The number density of sources (deg$^{-2}$) with flux density greater than
$S_{\lambda}$ is
\begin{eqnarray*}
\frac{dN(> S_{\lambda})}{d\Omega} &=& \int \frac{dN}{dV dt d\log M_b d\mu df_g}(M_b, \mu, f_g, z) \\
&& \tau(>S_{\lambda}, M_b, \mu, f_g, z) \frac{dV}{d\Omega dz}(z) d\log M_b d\mu df_g dz,
\end{eqnarray*}
where the term on the right-hand side first line is the comoving number density of mergers of
fixed most-massive progenitor baryonic mass $M_b$, merger mass ratio $\mu$, gas fraction at merger $f_g$,
and redshift $z$; $\tau(>S_{\lambda}, M_b, \mu, f_g, z)$ is the amount of time (duty cycle) a merger with given
$M_b$, $\mu$, and $f_g$ put at redshift $z$ has observed-frame flux density at wavelength $\lambda$
greater than $S_{\lambda}$; and $dV/d\Omega dz(z)$ is the comoving volume per square degree per unit redshift
at redshift $z$. We integrate over the redshift range 1 - 6
and assume a WMAP-7 cosmology \citep{Larson:2010}.

For each of our simulated galaxies we calculate the duty cycle for various sub-mm fluxes. We fit the
model duty cycles as a function of progenitor baryonic mass and merger mass ratio. We neglect any dependence
of the duty cycle on gas fraction or redshift, as all progenitor disks are initialized with $80\%$ gas (so that sufficient
gas remains at merger coalescence) and scaled to $z = 3$, and the negative $k$-correction causes the observed
sub-mm flux of a galaxy to be effectively independent of its redshift within the redshift range with which we are concerned.
Duty cycles in hand, we calculate the merger rates
using the semi-empirical model of \citet{Hopkins:2010merger_rates,Hopkins:2010IR_LF}. The model starts with an observed stellar
mass function (MF) \citep{Marchesini:2009}, assigns gas masses from observations, and places the galaxies in (sub)halos via a halo
occupation distribution
approach. Given halo-halo merger rates from $N$-body simulations, galaxy-galaxy merger rates are calculated by assuming the
galaxies merge on a dynamical friction timescale.

\section{Results}

\begin{figure}[!tb]
\plotone[angle=90,scale=0.5]{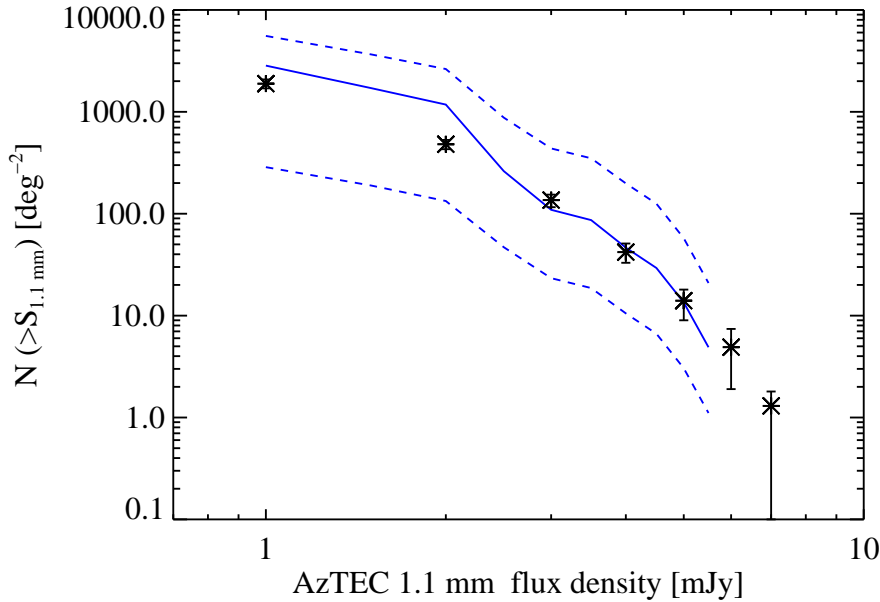}
\caption{AzTEC 1.1 mm number counts predicted by our fiducial model including the contributions from the initial infall,
first passage, and final coalescence of the two progenitor disks. We neglect the contribution from isolated disks as it is
sub-dominant for the flux range plotted here.
The ordinate is the cumulative number of sources per square degree with
1.1 mm flux density greater than the value of the abscissa. The solid line is the prediction calculated using the best-fit MF
of \citet{Marchesini:2009}. The dotted lines show the range of predictions when the 1$\sigma$ random uncertainties
and systematic uncertainties (from photometric redshifts and SED fitting)
from Marchesini et al. are used. Uncertainties on our calculated fluxes (owing primarily to the sub-resolution ISM treatment
and initial conditions of both the {\sc Gadget-2} simulations and the RT) are not included here but can have a significant effect
on the predicted counts. See \S\ref{S:discussion} for discussion.
\label{fig:num_cts}
}
\end{figure}

Figure \ref{fig:num_cts} shows the AzTEC 1.1 mm number counts from mergers predicted by our fiducial model, where
we have discarded all snapshots with gas fraction greater than 60\% when calculating the sub-mm duty cycles. The predictions
agree well with the \citet{Austermann:2010} SHADES data, which we use because to our knowledge this is the largest field
for which deep $\sim 1$ mm counts are available. Previous 850 \micron ~SCUBA counts were biased high by cosmic
variance \citep{Austermann:2010}; SCUBA-2 should remedy this problem.
One possible reason we do not have any galaxies with $S_{1.1 {\rm ~mm}} > 5.5$ mJy is because we do
not extrapolate beyond our largest simulated merger (a major merger of two disks with baryonic mass $4 \times 10^{11} M_{\odot}$),
and the $\sim 5$ sources with $S_{1.1 {\rm ~mm}} > 5.5$ mJy may be even more extreme, rare objects. Alternately, the discrepancy
may owe to our underprediction of the sub-mm flux at merger coalescence for the reasons outlined below, as a $\sim 10\%$
boost of the flux of our brightest sources is enough to account for all but the single brightest SMG observed in the SHADES field.

In our models isolated disks contribute negligibly to the `classical' SMG population ($S_{850 ~\mu {\rm m}} > 5$ mJy); a 60\%
gas disk with baryonic mass $4 \times 10^{11} M_{\odot}$ has $S_{850 ~\mu {\rm m}} \lesssim 4$ mJy
($S_{1.1 {\rm ~mm}} \lesssim 1.5$ mJy).
Mergers contribute to the SMG population in two ways: First, as has been suggested by numerous authors, some SMGs
are merger-driven starbursts. In this scenario, tidal torques at merger
coalescence induce a strong starburst and peak in AGN activity, both which cause the luminosity to sharply increase.
As a result, the dust luminosity and thus sub-mm flux increases, though this is partially mitigated by an increase
in the dust temperature. This is the scenario \citet{Narayanan:2010smg} studied in detail.
The second contribution is from the merging disks before they are strongly interacting.
The beam sizes of SCUBA and AzTEC are $\sim 30''$, which is $\sim 200$ kpc (comoving) at SMG redshifts ($z \sim 1-5$).
Thus as the massive, gas-rich, disks approach one another, both are contained in the beam for a significant time period.
Since individually the disks form stars rapidly and are dusty enough to have high -- but sub-SMG -- sub-mm fluxes, two in the
same $\sim 200$ kpc beam can yield a flux greater than the SMG threshold even if the disks' SFR is not elevated by
gravitational torques. While this mechanism relies on merging so that the two disks are in the same beam, the sub-mm
flux is not driven by the merger-induced starburst.
As shown in Figure \ref{fig:num_cts}, in our model the combination of these two mechanisms can account for the observed
number density of SMGs without invoking a non-standard IMF.

\section{Discussion} \label{S:discussion}

\begin{figure}[tb]
\plotone[angle=90,scale=0.5]{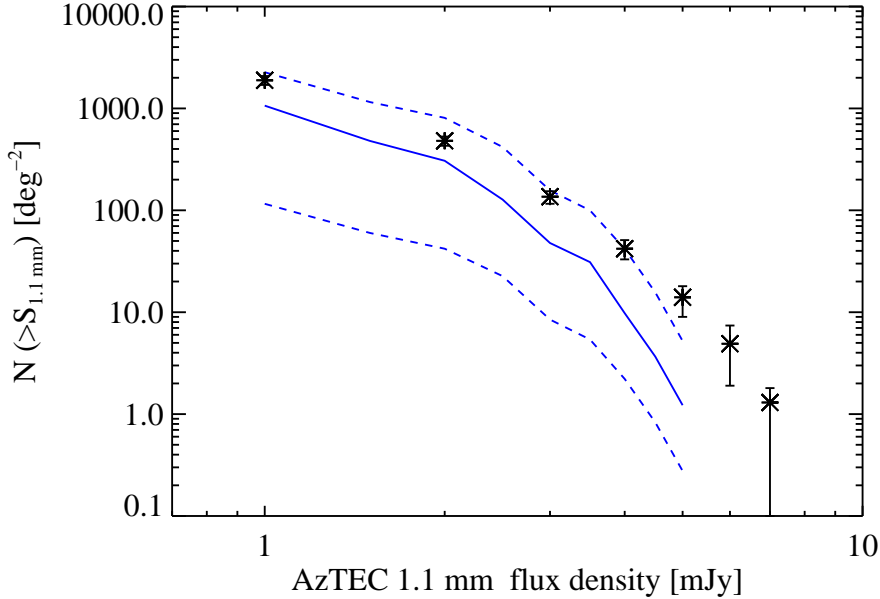}
\caption{Same as Figure \ref{fig:num_cts} but using only snapshots with gas fraction less than 0.4. This choice eliminates
most of the contribution from the infalling disks.\label{fig:num_cts_0.4}
}
\end{figure}

One primary advantage our simulations is that they have the resolution necessary to model the gravitationally-induced tidal
torques responsible for driving gas into the center of the merging galaxies and thus fueling the starburst and AGN. Furthermore,
the high-resolution simulations enable us to perform RT on a realistic 3-D distribution of stars, AGN,
and dust. We self-consistently calculate the thermal equilibrium temperature of each dust species including dust self-absorption.
These aspects of our models are essential for an accurate determination of the sub-mm flux. However, this accurate
modeling is very computationally expensive, and we must model a large parameter space. The number of high-resolution
models that must be run prevents us from performing cosmological simulations.
One significant limitation that results is that our simulations do not include cosmological gas accretion
\citep[e.g.,][]{Keres:2005,Keres:2009a}. Instead,
we initialize the disks with high gas fractions (80\%) so that a sufficient gas reservoir is present at the coalescence of the two galaxies.
Our somewhat crude treatment of the galaxies' gas content likely causes us
to underestimate the sub-mm flux at merger coalescence because the typical gas fractions at that time ($15-30$\%)
are less than the typical gas fractions observed ($\sim 40$\%) for both SMGs \citep{Tacconi:2006} and high-$z$ massive
star-forming galaxies of comparable mass \citep{Tacconi:2010}. Higher gas fraction can lead to a stronger starburst, greater
dust mass, and, consequently, higher sub-mm flux.
On the other hand, inclusion of disks with gas fractions higher than those observed can bias the counts upwards. Thus
our fiducial model may overpredict the contribution from the earliest stages of the merger when the two rapidly star-forming
disks are both in the beam but not yet strongly interacting.

To avoid including disks with unrealistically high gas fractions we have discarded all snapshots with gas
fraction greater than 60\% when calculating the counts plotted in Figure \ref{fig:num_cts}.
To check the sensitivity of our prediction to our gas fraction cut,
we also calculate the number counts using only snapshots with gas fraction less than 0.4. The results are shown in
Figure \ref{fig:num_cts_0.4}. The counts are less than observed by a factor of a few
for most bins when the best-fit Marchesini MF is used to compute the merger rates, but they are within the 1$\sigma$
upper limit except for the highest flux bin.

Despite our relatively sophisticated treatment of the RT, significant uncertainties in the sub-mm flux calculation
remain. One of the most important is the treatment of the sub-resolution ISM structure \citep{Younger:2009,Wuyts:2009}.
As described in \S\ref{S:methods},
we assume that the dust density is smooth on scales below the {\sc Gadget-2} resolution. We have used the total gas/metal density in
the {\sc Gadget-2} simulations to calculate the dust density, ignoring the SH03 multiphase breakdown.
An alternate treatment of the sub-resolution ISM is to discard the cold phase of the SH03 model, effectively assuming the cold
clouds have a negligible volume filling factor.
Observations suggest that in local ULIRGs the dense molecular gas component
of the ISM has a high volume filling factor \citep[e.g.,][]{Sakamoto:1999}, leading us to adopt the former treatment, but this
choice remains a source of uncertainty. To constrain this uncertainty, we have calculated the fluxes
for a restricted set of models where we have discarded the cold phase, which typically accounts for $\gtrsim 90\%$ of the dust mass.
Predictably, this leads to hotter SEDs, which results in lower sub-mm fluxes (typically by a factor of a few).
This provides unrealistically low attenuation when the attenuation from birth clouds is not included, but it is useful as a lower limit
on the sub-mm flux from a simulated galaxy.
Work to improve the {\sc Sunrise} treatment of the sub-resolution ISM structure is underway.

\section{Conclusion}

We have, for the first time, combined high-resolution 3-D hydrodynamic simulations of isolated and merging galaxies, radiative transfer,
and a semi-empirical merger rate model to predict number densities of SMGs. We find that isolated disks do not contribute
significantly to the classical SMG population, whereas mergers contribute both because of the coalescence-induced starburst
and because the two sub-mm bright but sub-SMG disks would not be resolved by the sub-mm instruments (e.g., SCUBA and AzTEC)
used to determine wide-field number counts. We constrain free parameters by observations whenever possible and carefully
test the effects of remaining uncertainties. We stress that we have not used a top-heavy IMF. Our predicted counts agree
reasonably well with observations, though significant modeling uncertainties remain. The agreement is due to a combination
of the relatively long sub-mm duty cycles of our simulated mergers \citep[as much as a few hundred Myr for the most massive mergers,
which is consistent with duty cycles estimated from observations;][]{Tacconi:2006}, the relatively higher number density of massive,
gas-rich mergers in our semi-empirical model, and the reduction in counts observed by recent
deep/wide surveys \citep{Austermann:2010} relative to previous results. Our ability to match
the observed counts without invoking a flat IMF indicates that, contrary to previous claims, observed SMG number densities do
not provide sufficient evidence to reject the null hypothesis of a universal IMF.

\bibliography{Hayward_C}

\begin{thebibliography}{}
\expandafter\ifx\csname natexlab\endcsname\relax\def\natexlab#1{#1}\fi
\expandafter\ifx\csname url\endcsname\relax
  \def\url#1{\texttt{#1}}\fi
\expandafter\ifx\csname urlprefix\endcsname\relax\def\urlprefix{URL }\fi
\providecommand{\eprint}[2][]{\url{#2}}

\bibitem[{Alexander et~al.(2005)Alexander, Bauer, Chapman, Smail, Blain,
  Brandt, \& Ivison}]{Alexander:2005}
Alexander, D.~M., Bauer, F.~E., Chapman, S.~C., Smail, I., Blain, A.~W.,
  Brandt, W.~N., \& Ivison, R.~J. 2005, \apj, 632, 736

\bibitem[{Austermann et~al.(2010)Austermann, Dunlop, Perera, Scott, Wilson,
  Aretxaga, Hughes, Almaini, Chapin, Chapman, Cirasuolo, Clements, Coppin,
  Dunne, Dye, Eales, Egami, Farrah, Ferrusca, Flynn, Haig, Halpern, Ibar,
  Ivison, van Kampen, Kang, Kim, Lacey, Lowenthal, Mauskopf, McLure, Mortier,
  Negrello, Oliver, Peacock, Pope, Rawlings, Rieke, Roseboom, Rowan-Robinson,
  Scott, Serjeant, Smail, Swinbank, Stevens, Velazquez, Wagg, \&
  Yun}]{Austermann:2010}
Austermann, J.~E., Dunlop, J.~S., Perera, T.~A., Scott, K.~S., Wilson, G.~W.,
  Aretxaga, I., Hughes, D.~H., Almaini, O., Chapin, E.~L., Chapman, S.~C.,
  Cirasuolo, M., Clements, D.~L., Coppin, K. E.~K., Dunne, L., Dye, S., Eales,
  S.~A., Egami, E., Farrah, D., Ferrusca, D., Flynn, S., Haig, D., Halpern, M.,
  Ibar, E., Ivison, R.~J., van Kampen, E., Kang, Y., Kim, S., Lacey, C.,
  Lowenthal, J.~D., Mauskopf, P.~D., McLure, R.~J., Mortier, A. M.~J.,
  Negrello, M., Oliver, S., Peacock, J.~A., Pope, A., Rawlings, S., Rieke, G.,
  Roseboom, I., Rowan-Robinson, M., Scott, D., Serjeant, S., Smail, I.,
  Swinbank, A.~M., Stevens, J.~A., Velazquez, M., Wagg, J., \& Yun, M.~S. 2010,
  \mnras, 401, 160

\bibitem[{Baugh et~al.(2005)Baugh, Lacey, Frenk, Granato, Silva, Bressan,
  Benson, \& Cole}]{Baugh:2005}
Baugh, C.~M., Lacey, C.~G., Frenk, C.~S., Granato, G.~L., Silva, L., Bressan,
  A., Benson, A.~J., \& Cole, S. 2005, \mnras, 356, 1191

\bibitem[{Blain et~al.(2002)Blain, Smail, Ivison, Kneib, \&
  Frayer}]{Blain:2002}
Blain, A.~W., Smail, I., Ivison, R.~J., Kneib, J.-P., \& Frayer, D.~T. 2002,
  Physics Reports, 369, 111

\bibitem[{Dav{\'e} et~al.(2010)Dav{\'e}, Finlator, Oppenheimer, Fardal, Katz,
  Kere{\v s}, \& Weinberg}]{Dave:2010}
Dav{\'e}, R., Finlator, K., Oppenheimer, B.~D., Fardal, M., Katz, N., Kere{\v
  s}, D., \& Weinberg, D.~H. 2010, \mnras, 404, 1355

\bibitem[{Groves et~al.(2008)Groves, Dopita, Sutherland, Kewley, Fischera,
  Leitherer, Brandl, \& van Breugel}]{Groves:2008}
Groves, B., Dopita, M.~A., Sutherland, R.~S., Kewley, L.~J., Fischera, J.,
  Leitherer, C., Brandl, B., \& van Breugel, W. 2008, \apjs, 176, 438

\bibitem[{Hopkins et~al.(2010{\natexlab{a}})Hopkins, Bundy, Croton, Hernquist,
  Keres, Khochfar, Stewart, Wetzel, \& Younger}]{Hopkins:2010merger_rates}
Hopkins, P.~F., Bundy, K., Croton, D., Hernquist, L., Keres, D., Khochfar, S.,
  Stewart, K., Wetzel, A., \& Younger, J.~D. 2010{\natexlab{a}}, \apj, 715, 202

\bibitem[{Hopkins et~al.(2010{\natexlab{b}})Hopkins, Younger, Hayward,
  Narayanan, \& Hernquist}]{Hopkins:2010IR_LF}
Hopkins, P.~F., Younger, J.~D., Hayward, C.~C., Narayanan, D., \& Hernquist, L.
  2010{\natexlab{b}}, \mnras, 402, 1693

\bibitem[{Jonsson(2006)}]{Jonsson:2006sunrise}
Jonsson, P. 2006, \mnras, 372, 2

\bibitem[{Jonsson et~al.(2010)Jonsson, Groves, \& Cox}]{Jonsson:2010sunrise}
Jonsson, P., Groves, B.~A., \& Cox, T.~J. 2010, \mnras, 403, 17

\bibitem[{Kennicutt(1998)}]{Kennicutt:1998}
Kennicutt, R.~C. 1998, \apj, 498, 541

\bibitem[{Kere{\v s} et~al.(2009)Kere{\v s}, Katz, Fardal, Dav{\'e}, \&
  Weinberg}]{Keres:2009a}
Kere{\v s}, D., Katz, N., Fardal, M., Dav{\'e}, R., \& Weinberg, D.~H. 2009,
  \mnras, 395, 160

\bibitem[{Kere{\v s} et~al.(2005)Kere{\v s}, Katz, Weinberg, \&
  Dav{\'e}}]{Keres:2005}
Kere{\v s}, D., Katz, N., Weinberg, D.~H., \& Dav{\'e}, R. 2005, \mnras, 363, 2

\bibitem[{Kov{\'a}cs et~al.(2006)Kov{\'a}cs, Chapman, Dowell, Blain, Ivison,
  Smail, \& Phillips}]{Kovacs:2006}
Kov{\'a}cs, A., Chapman, S.~C., Dowell, C.~D., Blain, A.~W., Ivison, R.~J.,
  Smail, I., \& Phillips, T.~G. 2006, \apj, 650, 592

\bibitem[{Kroupa(2001)}]{Kroupa:2001}
Kroupa, P. 2001, \mnras, 322, 231

\bibitem[{Larson et~al.(2010)Larson, Dunkley, Hinshaw, Komatsu, Nolta, Bennett,
  Gold, Halpern, Hill, Jarosik, Kogut, Limon, Meyer, Odegard, Page, Smith,
  Spergel, Tucker, Weiland, Wollack, \& Wright}]{Larson:2010}
Larson, D., Dunkley, J., Hinshaw, G., Komatsu, E., Nolta, M.~R., Bennett,
  C.~L., Gold, B., Halpern, M., Hill, R.~S., Jarosik, N., Kogut, A., Limon, M.,
  Meyer, S.~S., Odegard, N., Page, L., Smith, K.~M., Spergel, D.~N., Tucker,
  G.~S., Weiland, J.~L., Wollack, E., \& Wright, E.~L. 2010, ArXiv e-prints.
  \eprint{1001.4635}

\bibitem[{{Lu} et~al.(2010){Lu}, {Mo}, {Weinberg}, \& {Katz}}]{Lu:2010}
{Lu}, Y., {Mo}, H.~J., {Weinberg}, M.~D., \& {Katz}, N.~S. 2010, ArXiv
  e-prints. \eprint{1004.2518}

\bibitem[{Marchesini et~al.(2009)Marchesini, van Dokkum, {F{\"o}rster
  Schreiber}, Franx, Labb{\'e}, \& Wuyts}]{Marchesini:2009}
Marchesini, D., van Dokkum, P.~G., {F{\"o}rster Schreiber}, N.~M., Franx, M.,
  Labb{\'e}, I., \& Wuyts, S. 2009, \apj, 701, 1765

\bibitem[{Narayanan et~al.(2010{\natexlab{a}})Narayanan, Cox, Hayward, \&
  Hernquist}]{Narayanan:2010ks}
Narayanan, D., Cox, T.~J., Hayward, C.~C., \& Hernquist, L. 2010{\natexlab{a}},
  ArXiv e-prints. \eprint{1005.3020}

\bibitem[{Narayanan et~al.(2009)Narayanan, Cox, Hayward, Younger, \&
  Hernquist}]{Narayanan:2009}
Narayanan, D., Cox, T.~J., Hayward, C.~C., Younger, J.~D., \& Hernquist, L.
  2009, \mnras, 400, 1919

\bibitem[{Narayanan et~al.(2008)Narayanan, Cox, Kelly, Dav{\'e}, Hernquist,
  Matteo, Hopkins, Kulesa, Robertson, \& Walker}]{Narayanan:2008}
Narayanan, D., Cox, T.~J., Kelly, B., Dav{\'e}, R., Hernquist, L., Matteo,
  T.~D., Hopkins, P.~F., Kulesa, C., Robertson, B., \& Walker, C.~K. 2008,
  \apjs, 176, 331

\bibitem[{Narayanan et~al.(2010{\natexlab{b}})Narayanan, Dey, Hayward, Cox,
  Bussmann, Brodwin, Jonsson, Hopkins, Groves, Younger, \&
  Hernquist}]{Narayanan:2010dog}
Narayanan, D., Dey, A., Hayward, C.~C., Cox, T.~J., Bussmann, R.~S., Brodwin,
  M., Jonsson, P., Hopkins, P.~F., Groves, B., Younger, J.~D., \& Hernquist, L.
  2010{\natexlab{b}}, ArXiv e-prints. \eprint{0910.2234}

\bibitem[{Narayanan et~al.(2010{\natexlab{c}})Narayanan, Hayward, Cox,
  Hernquist, Jonsson, Younger, \& Groves}]{Narayanan:2010smg}
Narayanan, D., Hayward, C.~C., Cox, T.~J., Hernquist, L., Jonsson, P., Younger,
  J.~D., \& Groves, B. 2010{\natexlab{c}}, \mnras, 401, 1613

\bibitem[{Sakamoto et~al.(1999)Sakamoto, Scoville, Yun, Crosas, Genzel, \&
  Tacconi}]{Sakamoto:1999}
Sakamoto, K., Scoville, N.~Z., Yun, M.~S., Crosas, M., Genzel, R., \& Tacconi,
  L.~J. 1999, \apj, 514, 68

\bibitem[{Schmidt(1959)}]{Schmidt:1959}
Schmidt, M. 1959, \apj, 129, 243

\bibitem[{Springel(2005)}]{Springel:2005gadget}
Springel, V. 2005, \mnras, 364, 1105

\bibitem[{Springel \& Hernquist(2003)}]{Springel:2003}
Springel, V., \& Hernquist, L. 2003, \mnras, 339, 289

\bibitem[{Springel et~al.(2005)Springel, Matteo, \&
  Hernquist}]{Springel:2005feedback}
Springel, V., Matteo, T.~D., \& Hernquist, L. 2005, \mnras, 361, 776

\bibitem[{Tacconi et~al.(2010)Tacconi, Genzel, Neri, Cox, Cooper, Shapiro,
  Bolatto, Bouch{\'e}, Bournaud, Burkert, Combes, Comerford, Davis,
  {F{\"o}rster Schreiber}, Garcia-Burillo, Gracia-Carpio, Lutz, Naab, Omont,
  Shapley, Sternberg, \& Weiner}]{Tacconi:2010}
Tacconi, L.~J., Genzel, R., Neri, R., Cox, P., Cooper, M.~C., Shapiro, K.,
  Bolatto, A., Bouch{\'e}, N., Bournaud, F., Burkert, A., Combes, F.,
  Comerford, J., Davis, M., {F{\"o}rster Schreiber}, N.~M., Garcia-Burillo, S.,
  Gracia-Carpio, J., Lutz, D., Naab, T., Omont, A., Shapley, A., Sternberg, A.,
  \& Weiner, B. 2010, Nature, 463, 781

\bibitem[{Tacconi et~al.(2006)Tacconi, Neri, Chapman, Genzel, Smail, Ivison,
  Bertoldi, Blain, Cox, Greve, \& Omont}]{Tacconi:2006}
Tacconi, L.~J., Neri, R., Chapman, S.~C., Genzel, R., Smail, I., Ivison, R.~J.,
  Bertoldi, F., Blain, A., Cox, P., Greve, T., \& Omont, A. 2006, \apj, 640,
  228

\bibitem[{Wuyts et~al.(2009)Wuyts, Franx, Cox, {F{\"o}rster Schreiber},
  Hayward, Hernquist, Hopkins, Labb{\'e}, Marchesini, Robertson, Toft, \& van
  Dokkum}]{Wuyts:2009}
Wuyts, S., Franx, M., Cox, T.~J., {F{\"o}rster Schreiber}, N.~M., Hayward,
  C.~C., Hernquist, L., Hopkins, P.~F., Labb{\'e}, I., Marchesini, D.,
  Robertson, B.~E., Toft, S., \& van Dokkum, P.~G. 2009, \apj, 700, 799

\bibitem[{Younger et~al.(2009)Younger, Hayward, Narayanan, Cox, Hernquist, \&
  Jonsson}]{Younger:2009}
Younger, J.~D., Hayward, C.~C., Narayanan, D., Cox, T.~J., Hernquist, L., \&
  Jonsson, P. 2009, \mnras, 396, L66

\end{thebibliography}

\end{document}